\documentclass[acmsmall]{acmart}

\AtBeginDocument{%
  \providecommand\BibTeX{{%
    \normalfont B\kern-0.5em{\scshape i\kern-0.25em b}\kern-0.8em\TeX}}}

\setcopyright{acmcopyright}
\acmJournal{PACMHCI}
\acmYear{2022} \acmVolume{6} \acmNumber{CSCW1} \acmArticle{48} \acmMonth{4} \acmPrice{15.00}\acmDOI{10.1145/3512895}

\received{July 2021}
\received[revised]{November 2021}
\received[accepted]{November 2021}

\begin{document}

\title[A ``Distance Matters'' Paradox]{A ``Distance Matters'' Paradox: Facilitating Intra-Team Collaboration Can Harm Inter-Team Collaboration}

\author{Xinlan Emily Hu}
\email{xehu@wharton.upenn.edu}
\orcid{0000-0001-9439-3498}
\affiliation{
\institution{The Wharton School, University of Pennsylvania}
\country{U.S.A.}
}

\author{Rebecca Hinds}
\email{rhinds@stanford.edu}
\orcid{0000-0003-4849-8162}
\affiliation{
\institution{Stanford University}
\country{U.S.A.}
}

\author{Melissa A. Valentine}
\email{mav@stanford.edu}
\orcid{0000-0001-7517-4054}
\affiliation{
\institution{Stanford University}
\country{U.S.A.}
}

\author{Michael S. Bernstein}
\email{msb@cs.stanford.edu}
\orcid{0000-0001-8020-9434}
\affiliation{
 \institution{Stanford University}
  \country{U.S.A.}
 }

\renewcommand{\shortauthors}{Anonymous}

\begin{abstract}
By identifying the socio-technical conditions required for teams to work effectively remotely, the Distance Matters framework has been influential in CSCW since its introduction in 2000. Advances in collaboration technology and practices have since brought teams increasingly closer to achieving these conditions. This paper presents a ten-month ethnography in a remote organization, where we observed that despite exhibiting excellent remote collaboration, teams paradoxically struggled to collaborate across team boundaries. We extend the Distance Matters framework to account for inter-team collaboration, arguing that challenges analogous to those in the original intra-team framework --- common ground, collaboration readiness, collaboration technology readiness, and coupling of work --- persist but are actualized differently at the inter-team scale. Finally, we identify a fundamental tension between the intra- and inter-team layers: the collaboration technology and practices that help individual teams thrive (e.g., adopting customized collaboration software) can also prompt collaboration challenges in the inter-team layer, and conversely the technology and practices that facilitate inter-team collaboration (e.g., strong centralized IT organizations) can harm practices at the intra-team layer. The addition of the inter-team layer to the Distance Matters framework opens new opportunities for CSCW, where balancing the tension between team and organizational collaboration needs will be a critical technological, operational, and organizational challenge for remote work in the coming decades.
\end{abstract}

\ccsdesc[500]{Human-centered computing~Computer supported cooperative work}

\keywords{distance, teams, workplace, distributed work, remote work, future of work, ethnography, collaboration technology}

\maketitle

\section{Introduction}
Distance Matters is a landmark framework~\cite{olson2000distance} and research agenda~\cite{olson2008theory, olson2013working, bjorn2014does} in Computer-Supported Cooperative Work. It describes socio-technical conditions inherent in distance work that present challenges for collaboration, and argues that, in order to position themselves for success, groups must exhibit high common ground and loosely coupled work, as well as readiness for both collaboration and collaboration technology. The framework draws from studies comparing a distributed team to a colocated counterpart: in the conference rooms of colocated teams,~\citeauthor{olson2000distance} observed participants wheeling their chairs across the room to form dynamic clusters. Team members explained complicated ideas simply by gesturing at an imaginary whiteboard in the air. That conference room, and the seeming effortlessness of collaboration within it, constituted what~\citeauthor{olson2000distance} argued that groups working remotely had yet to replicate.

Over the years, however, improved technology, ubiquitous adoption of collaboration tools, and well-established usage norms have helped groups reshape their prospects of recreating a virtual conference room.~\citet{bjorn2014does}, for instance, applied the framework to four empirical cases in which work groups collaborated at a distance as a matter of course. Their study argues that changing workplace conditions, including improved collaboration technology readiness, call the salience of several themes in the original framework into question. Recent research has also found that geographic and temporal distance matter far less than other factors, such as how far apart collaborators are in an organization's hierarchy, and how different the team members' roles are~\cite{wang2021organizational}.~\citet{warshaw2016distance} and others~\cite{leonardi2010connectivity, prikladnicki2016virtual} suggest that we can now expect more teams to succeed while working remotely --- almost to the point of asking whether distance still matters. Even Judith Olson has stated, in a \textit{New York Times} article in June~2021, that distance now matters less: ``because of the technology these days, we're actually inching closer and closer to replicating the office''~\cite{millernyt2021}.

To understand the current salience of distance in remote teams, we performed a ten-month ethnography of remote teams at an approximately 2000-person government lab, which had primarily remote operations during the year of our study. We observed that remote teams had indeed become quite fluent in collaborating at a distance, but that they paradoxically struggled in \textit{inter-team} (between team) collaboration. In one case, for instance, teams chose software that was best suited to their local needs, but the software choice caused significant coordination issues among external groups. One cross-functional collaborator on the business support team found herself using nine collaboration tools when she only needed the functionality of two; another, tasked with communicating with employees across multiple divisions, struggled to maneuver across a morass of communication platforms.

We draw on our ethnographic results, and on both CSCW and organizational research on inter-team collaboration (e.g.,~\cite{orlikowski1995learning,grudin1994groupware}), to extend the Distance Matters framework. Our extended framework accounts for the additional distance work demands introduced through inter-team collaboration. In our analysis, the same themes --- common ground, collaboration readiness, collaboration technology readiness, and coupling of work --- continue to apply, but are actualized differently at this new scale, leading to novel design challenges. For instance, while collaboration technology is now adept at creating interactive virtual team workspaces, complete with whiteboards, real-time communication, and a sense of social presence, most of the technologies used by today's remote teams have not been designed to facilitate the same sense of presence and common ground \textit{between teams}. When a user participates in a standup via Slack with their core team, they might gain shared understanding of the status of their own team's work, but they may not understand the status of other teams' work and, for example, whether a shared cross-functional deliverable is on schedule. Collaboration technology replicates the conference room, but not the office buildings, facilities, and complex network of connected locations that compose multi-functional, global organizations.

The intra- and inter-team layers of the Distance Matters framework exist in fundamental tension. Just as collaboration software can suffer if it empowers one group over another~\cite{grudin1994groupware}, and just as different teams and groups develop distinctive norms, meanings, practices and cultures when enacting the same technologies~\cite{orlikowskITresearch}, comparing the two layers of the Distance Matters framework calls out how solutions that are optimal at one layer can cause problems at the other. For example, adopting specialized collaboration tools aids the team, but makes collaboration outside the team more difficult; creating a powerful centralized IT organization helps support shared needs but stymies collaborative practices within the team.

Understanding the Distance Matters themes holistically, at both the intra- and inter-team level, is therefore critical to designing a distributed workplace that can effectively accommodate multi-team organizations. The tightrope of balancing collaboration needs between teams and organizations will only become more central as organizations adopt permanent flexible work arrangements~\cite{neeley2021remote} or do away with the office building altogether~\cite{rhymer2020location}. Resolving this tension, and designing for the inter-team level without sacrificing the intra-team level (and vice versa), requires a complementary perspective on a classic frontier of exploration for CSCW. This challenge creates opportunities to design technologies to facilitate inter-team collaboration, establish collaboration structures that facilitate teams of teams~\cite{salas2008teams}, and create new roles that leverage technology to bridge inter-team gaps. Ultimately, this research aims to refocus the lens of a central CSCW theory to help designers and researchers make sense of a growing source of collaboration breakdowns in remote work.

In the next section, we review the Distance Matters framework and the research it inspired, as well as the research on inter-team collaboration that motivates our question --- whether improved and ubiquitous technology has at last erased the perils of working at a distance. We then describe the methodology of our study and our extension of the framework. We conclude with a discussion of practical implications.

\section{Related Work} 
This section proceeds in three parts. We first review the Distance Matters framework and the research it inspired, particularly with respect to technology requirements for remote team collaboration. In the second part, we note that these technology requirements are rooted in an organizational context, and we review CSCW literature discussing organizational tensions over shared technology. Finally, we draw from organizational theory to elaborate on why such tensions might emerge, motivating our extension of the ``Distance Matters'' framework.

\subsection{Distance Matters and the Remote Team}\label{dm-rw}
Focused primarily on teams, the Distance Matters literature has developed rich theories for both the advantages and drawbacks of working at a distance. The original paper by~\citet{olson2000distance} has a vast, interdisciplinary reach; its 2,600+ citations span computer science, psychology, and management, among other fields. We therefore focus on this framework because it is a confluence of the remote work conversation across multiple disciplines --- and a crucial step to bridging the gap between intra-team and inter-team collaboration.

\begin{table}[h!]
\begin{tabular}{@{} p{0.2\textwidth} | p{0.7\textwidth} @{}}
\toprule
 Common Ground & ``Knowledge that the participants have in common, and they are aware that they have it in common'' (p. 157). \\[0.5cm]
 Collaboration Readiness & ``A willingness to share…[that] aligns with the incentive structure'' (p. 164). \\[0.5cm]
 Collaboration Technology Readiness & The ``habits and infrastructure'' around technology, including ``alignment of technology support, existing patterns of everyday usage, and the requirements for a new technology'' (p. 164-5). \\[0.5cm]
 Coupling of Work & ``The extent and kind of communication required by the work'' (p. 162). \\[0.5cm]
 Organizational Managerial Aspects & ``What the manager can do to ensure that the collaboration is successful,'' including aligning goals, designing reward structures, planning, and communicating decisions (p. 49~\cite{olson2013working}). \\
 \bottomrule
\end{tabular}
 \caption{The Original Distance Matters concepts. The first four rows originate from~\citet{olson2000distance}, while the last row originates from~\citet{olson2013working} .}
 \label{tab:dm-originalconcepts}
\end{table}

The original Distance Matters paper defines its four central concepts as shown in the first four rows of  Table~\ref{tab:dm-originalconcepts}. Later work~\cite{olson2013working} defines a separate category for management activities (row 5), in which leaders should engage in practices such as aligning goals, designing reward structures, creating explicit plans, and communicating decisions.

The themes from the framework have continued to ignite conversation. Subsequent work has explored the various dimensions of distance work in greater detail, including the potential for distanced work to engender greater conflict~\cite{hinds2003out, mannix2002phenomenology, hinds2005understanding, cramton2002attribution}; the effects of shared identity~\cite{nardi2002place, bos2010shared}; and different types of common ground~\cite{mao2019data}. Multiple papers share the title of “Does Distance Still Matter”~\cite{wolf2008does, bjorn2014does, jolak2018does}, with some concluding that it does matter~\cite{jolak2018does, jolak2020design}, some that it now matters less~\cite{wolf2008does, prikladnicki2016virtual} (including a quote from Judith Olson in~\cite{millernyt2021}), and some arguing that certain aspects now matter more than others~\cite{bjorn2014does}.

More broadly, the implicit Holy Grail is for a remote team to technologically replicate (or do even better than, vis-à-vis \textit{beyond being there}~\cite{hollan1992beyond}) the low-friction and high-frequency interactions of being in person.~\citet{kraut2002understanding} describes the key mechanisms through which proximity benefits collaboration, such as visibility, copresence, and cotemporality. Conversely, perceived geographic distance decreases one's willingness to cooperate with partners~\cite{bradner2002distance}, and those whose identities are closely tied to interpersonal relationships may see the salience of their relationships decrease when interacting over technology~\cite{lee2015making}. These findings have motivated a long research history in CSCW focused on creating greater social awareness in remote collaboration, with numerous applications designed to create a shared context and sense of presence (see reviews by~\citet{gross2005user} and by~\citet{olson2012collaboration}).

The ability for technology to sufficiently convey a presence-like richness is often the determining factor for whether distance matters or not. When studies find that distance still matters, the grievance is that technology remains a “pale imitation of face-to-face interaction”~\cite{jolak2018does}; when studies find that distance now matters less, the reason is that teams are now capable of using technology more effectively~\cite{wolf2008does, bjorn2014does, leonardi2010connectivity, millernyt2021}. Even when the task and technology do not initially align, teams adapt to their tools over time, effectively ``customizing'' them through a social process~\cite{fuller2009does}. This may take the form of consciously connecting people and artifacts (see \textit{relation work}~\cite{bjorn2011relation}), or embedding shared understanding into a virtual object (see \textit{Socially-Implicated Work Objects}~\cite{muller2017stick}). In general, the literature emphasizes that close coupling between the team and technology is critical to a successful collaboration.

While the Distance Matters framework has clearly been very influential, the majority of the literature focuses on teams as a unit of analysis. The emphasis on teams is highlighted in the review paper by~\citet{raghuram2019virtual}, in which most papers citing Distance Matters are grouped in the “Virtual Teams” cluster. The notion of recreating face-to-face interactions --- like the original framework --- is inspired by a metaphor of small group collaboration. As entire organizations shift to working remotely, the problems, opportunities, and design challenges related to remote collaboration shift with them. Our research extends this prior literature by articulating the Distance Matters framework as it explains collaboration across remote team boundaries.

\subsection{Seeds of Tension in the Remote Organization}\label{cscw-org-rw}
Next, we review the organizational perspective in CSCW, which we draw on to expand beyond the lens of individual teams. The organizational perspective is critical in CSCW, as increasingly, organizations are dependent on~\cite{olson1997research}, and even \textit{created} by~\cite{valentine2017flash}, collaboration technology systems. In their~\citeyear{grudin1997organizational} paper,~\citet{grudin1997organizational} describe four phases of implementing technology: Initiation, Acquisition, Implementation \& Use, and Impacts \& Consequences. Each of these phases can surface tensions between the high-level goals of the technology and the needs of on-the-ground users. For example, a decision to purchase a particular piece of software at the Initiation phase may not align with organizational needs, thus leading to poor productivity-related outcomes. On the other hand, a decision \textit{not} to support a particular software can also be troublesome and ``create difficulties later for activities that cross the boundaries between work units'' (p. 1464). In their ethnography of the specialized biology software Worm Community System (WCS),~\citet{leigh1994steps} found certain ``double binds'' in technology infrastructure when using collaborative software across team boundaries. Something intuitive to one team --- for example, using a terminal command --- may be difficult for another, leading to resource allocation conflicts.

A common thread in CSCW's studies of organizations is the interrelationship between the organization and the technology, and the challenges that may exist across roles and boundaries (for example, when a manager decides which tools their employees may use, or when one team's software choice impacts another). These tensions are rooted in organizational theory related to inter-team collaboration. In the next section, we present a more comprehensive review of this interdisciplinary set of literature and motivate our extension of the Distance Matters framework by connecting the threads from Sections~\ref{dm-rw} to~\ref{org-rw}.  

\subsection{Organizational Theory and the Extended Framework}\label{org-rw}
Four major themes from organizational theory reveal why inter-team collaboration differs from intra-team collaboration. First, team members who interact on a regular basis and solve similar problems together develop similar mental models and shared language~\cite{cannonbowersetal,lewisgroupcognition,rentschgreatminds}. These ongoing, repeat interactions around a shared problem become shared “thought worlds” within the team, but also become interpretive barriers between teams~\cite{doughertyinterpretive}. Interdependent teams may use different words to refer to the same concepts, creating translation issues at the boundaries~\cite{boland1995perspective, waring2018information}. This well-established dynamic highlights how team members’ common ground understanding within the team influences their collaborations across the team boundary~\cite{anconacaldwell, joshi}. 

Second, the structure of interdependence between teams can be more complicated than the structure of interdependence within teams. The interdependent relationships among teams can sometimes follow the structure of an organization’s products or services, setting up complex interrelated networks~\cite{henderson,sosamisalignment}. These networks become hard to visualize, model, or understand~\cite{huising,tuckerhospitals}, so designing collaboration or intervening to improve collaboration within these organizational systems is quite complex~\cite{barrettetal, mavasq}. Moreover, social dynamics such as status hierarchies and power differences develop within these complex relationships, which create additional challenges for collaboration~\cite{contuetal,metiuowningcode,pfefferorganizaitonal,pfefferexternalcontrol}. This second theme thus calls into question what collaboration readiness looks like between teams in such complex relational networks, where clear incentive alignment seems difficult to accomplish.

Third, and relatedly, different teams, groups, and occupations develop different norms, meanings, practices, and cultures when enacting technologies in their local work~\cite{orlikowskITresearch}. These different meanings become consequential when the different parties interact to transform the meanings of the technologies at their boundaries~\cite{bechkyobjectlessons,bechkysharingmeaning}. Groups and teams seek to ensure their own interests, even as they use dynamic coordination technologies to collaborate with their interdependent partners~\cite{kelloggtrading}. 

Finally, an emerging theme in organizations research involves the consistent finding that team membership is much messier and more complex than prior research models have accounted for~\cite{kerrisseyintothefray,wagemanecology}. People often belong to many teams and groups at the same time, even those with competing goals~\cite{olearyteams}. Members move dynamically between some of these different thought worlds and technology-sub-cultures described above~\cite{croninetal,hackmangroupbehavior,larsonleadingteams,mortensonteam}. It is not clear what the design opportunities and challenges are for the complex group membership practices that undergird modern organizations.

These research findings raise the question of what collaboration technology readiness means when it involves interdependent teams whose interests are not perfectly aligned. Past research suggests that the technology for inter-team collaboration may be less focused on awareness or copresence (trying to reproduce interactions in a conference room) and may need to be better configured for managing boundaries between specialized teams. However, these technological requirements can come into conflict with the requirement of copresence at the inter-team level, described in Section~\ref{dm-rw}.

In sum, an overarching challenge for the socio-technical conditions known to enable intra-team collaboration is that teams can ``over-optimize'' their local environments to the detriment of the larger organization; as~\citet{dechurch2010perspectives} point out, “teams can be too cohesive,” such that team processes detract from system-level ones. ``One cannot afford to ignore this reality,'' they warn, ``and assume that building successful teams will translate into successful systems of teams'' (p. 331). To explore whether and how distance matters differently for inter-team collaboration, we conducted an ethnography of a remote organization that has complex, multi-functional operations. We present empirical evidence for how the differences between the two levels becomes salient. We then discuss how these insights contribute to research at the intersection of collaboration and technology, enabling a reconfiguration of the Distance Matters framework.

\section{Methods}
\begin{figure}
 \includegraphics[width = 0.9\textwidth]{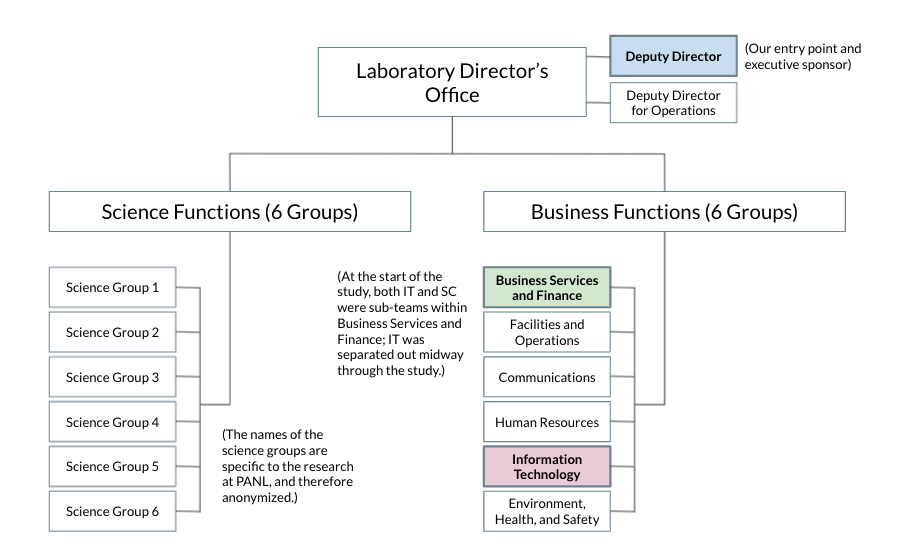}
 \caption{PANL's Organizational chart. The laboratory consists of six groups performing scientific functions, and six performing business functions. The science groups are not labeled, since their names reference research unique to PANL, and can therefore be used to identify the organization. Business Services and Finance, which is highlighted in green, originally housed both of the teams selected for our case study (IT and SC); IT was later reorganized as a separate group (highlighted in light purple). The deputy director of the laboratory, highlighted in blue, was our primary entry point to the field site and supported the research.}
 \label{fig:panl-org-chart}
\end{figure}

The setting for our study is (pseudonymously) the Particle Accelerator National Laboratory (PANL), a Department of Energy-funded National Laboratory in the United States. PANL is noted for research in the natural sciences conducted on its particle accelerator. The lab has approximately 2000 full-time employees, divided into multiple sub-organizations. These sub-organizations are split along scientific (e.g., physical sciences) and business (e.g., Human Resources, Information Technology) functions. Figure~\ref{fig:panl-org-chart} illustrates the organizational hierarchy at PANL.

We conducted a mixed-method qualitative study, with data originating from three sources: ethnographic fieldwork, semi-structured interviews, and a survey. Materials for the interview and survey are included in the appendix. Our observations took place over ten months between September 2020 and July 2021, a period during which nearly all of PANL's physical operations transitioned to remote work due to the COVID-19 pandemic.

\subsection{Field Site Access}
Our entry point for the study was a senior director of the laboratory, who was interested in collecting insights about PANL's adaptation to remote work. An acquaintance connected our research team with the senior director, who then gathered a cross-functional task force of PANL employees to work closely with the research team. The PANL task force sponsored the research by providing access to data, recruiting participants for interviews, disseminating surveys, and inviting the researchers to join relevant meetings and town hall events.

\subsection{Ethnographic Data Collection}
Ethnographic studies provide the methodological advantage of producing real-time data during a period of change~\cite{kellogg2009operating}. Because we were interested in the period of change as PANL adapted to remote work, ethnography was particularly appropriate. For this study, we used methods informed by a previous history of ethnographic fieldwork in CSCW (see reviews by~\citet{mcdonald2019reliability},~\citet{hughes1994moving}, and others~\cite{le2008view, millen2000rapid, rouncefield1994working}).

\subsubsection{Selection of Cases}
To understand the coordination challenges of working remotely across multiple teams, we selected two multi-team systems as case studies. Each one bridged diverse backgrounds and functions.

The first case was the Central Information Technology (IT) team, the unit in charge of data storage, hardware and software, networking, and other important technology functions. Central IT was identified as a key function in PANL’s future of work efforts; the group was a central node in a vast array of interactions, and regularly interfaced with the rest of the laboratory, including most scientific functions.

The second case was the Supply Chain Management (SC) team, responsible for procuring materials for other functions at PANL. Like Central IT, SC interfaced with a diverse set of teams across the laboratory. Both SC and IT were seen as critical ``support functions'' that could be completed remotely, and were therefore of interest to the study. Until midway through the study, both teams were also in the same sub-organization; IT was later separated, as shown in Figure~\ref{fig:panl-org-chart}.

Additionally, both teams exhibited prior remote work experience. The IT team members, due to their training and expertise, were unsurprisingly a technologically-savvy group. Although most IT employees worked on-site at PANL before the pandemic, several senior members of the team were fully remote before the pandemic, and therefore the group had established a successful remote work infrastructure.

Pre-pandemic, most of the SC team members worked from home at least two days per week. SC team members were extremely invested in the team's collaboration processes, and it held regular training sessions for customers to understand its functions. SC also maintained an annual scorecard measuring its customer success based on internal surveys. The team attained scores of 82/100, 91/100, and 87/100 in the three years preceding the pandemic (scores greater than or equal to 82 were considered ``Excellent'').

\begin{figure}
 \includegraphics[width = 0.9\textwidth]{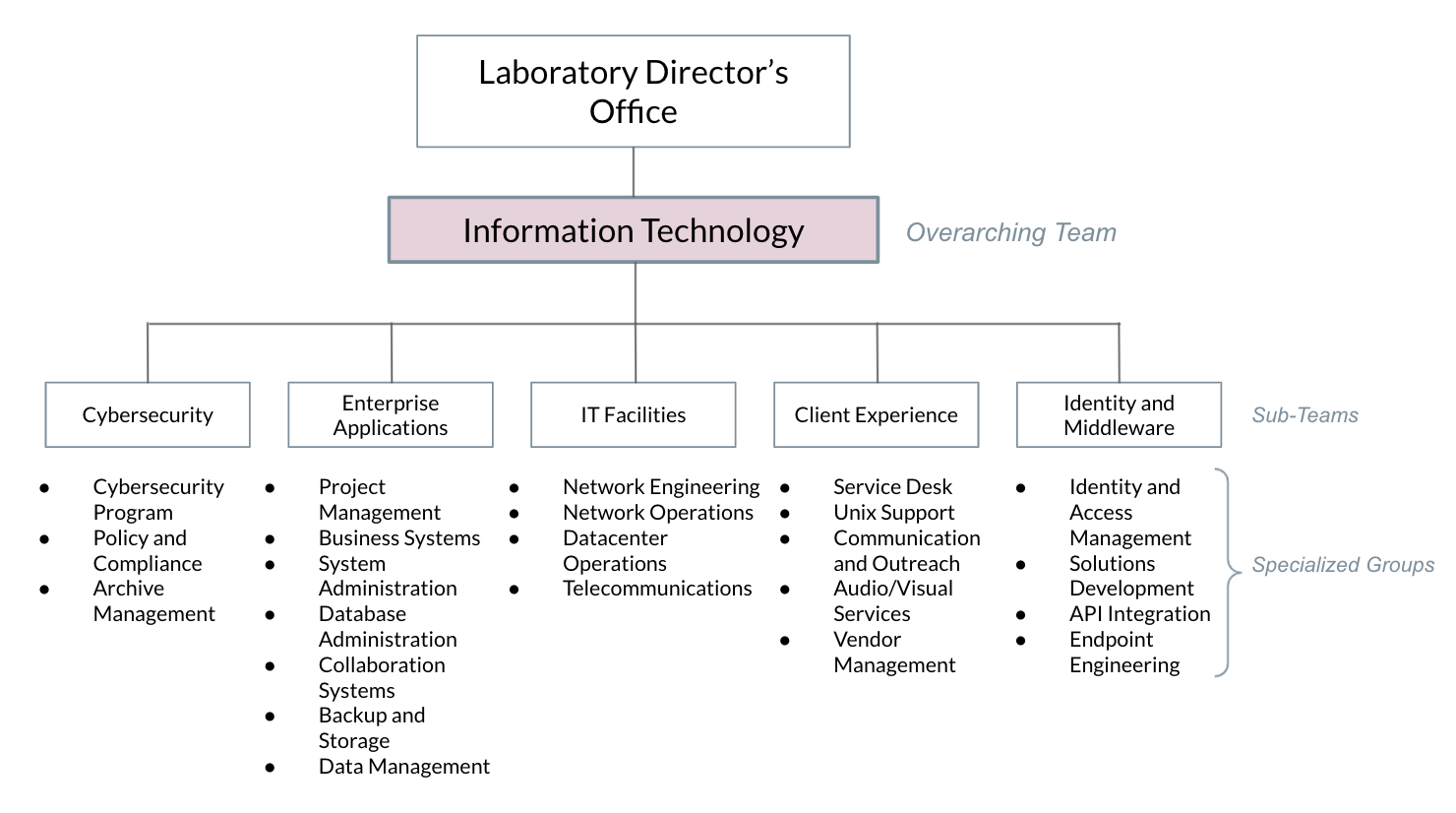}
 \caption{The IT Group's Organizational chart, used to illustrate the terminology of teams, sub-teams, and specialized groups.}
 \label{fig:it-org-chart}
\end{figure}

\subsubsection{Terminology of Teams}
We describe both IT and SC as ``teams,'' even though each team comprised dozens of employees and multiple constituent sub-groups. Traditionally, there is no exact specification of a team's size; the number of members has ranged from two to eight for a small group~\cite{hackman1970effects,bakeman1974size,hare1976handbook}, to 32 in a laboratory experiment~\cite{mao2016experimental}, to industry teams of over 300 members~\cite{cusumano1997microsoft}. Previous research has also examined scaffolds for organizing teams of 200 members~\cite{scerri2004scaling}.

For clarity in this paper, we adopt the following terminology, building on~\citet{scerri2004scaling}'s language of \textit{teams} and \textit{sub-teams}. We also use the IT group's organizational chart (Figure~\ref{fig:it-org-chart}) to illustrate the relevant concepts.
\begin{itemize}
    \item A \textit{team} is the overarching group of individuals who work in the same discipline, similar to how industry teams are defined as all individuals working on a single product~\cite{cusumano1997microsoft}. Each of the IT and SC groups is single team. This terminology also aligns with the way in which collaborators referred to IT and SC; that is, a scientist would refer to working with IT as a single team or entity, rather than to a specific sub-team within IT.
    \item A \textit{sub-team} is a group within a team tasked with handling a variety of tasks within a specific area; as illustrated in Figure~\ref{fig:it-org-chart}, the IT team is divided into five sub-teams. Each consists of between 10-30 individuals.
    \item A \textit{specialized group} consists of individuals who work to perform a specialized function. Most specialized groups consist of 3-8 members, although some functions (e.g., Service Desk) have 15, and other functions (e.g., Telecommunications) have just two members. Note, therefore, that specialized groups are not necessarily smaller than sub-teams; the distinction lies in hierarchical structure rather than group size.
    \item \textit{Inter-team collaboration} occurs when members of different teams must work towards a shared purpose. When SC works with chemists to order materials, for instance, this constitutes inter-team collaboration.
    \item Similarly, \textit{intra-team collaboration} occurs when sub-teams within a team interact with one another: for instance, the Cybersecurity sub-team may need to collaborate with the Identity and Middleware sub-team to develop a more secure login system.
\end{itemize}

\subsubsection{Ethnographic Data}
To collect ethnographic data, we engaged as participant observers at PANL while collecting data to inform senior leaders' decisions related to transitioning to remote work. This access permitted close observations and detailed field notes of both formal meetings and informal interactions. Researchers' notes captured both the information shared aloud during Zoom meetings and information shared via the chat --- which enabled us to document informal conversation and banter in more detail than would have likely been gleaned via a comparable in-person meeting. We also documented emails and messages exchanged with PANL staff, and we collected extensive archival data on both daily work and planned changes. The vast majority of data, including meeting observations, was collected online, via Zoom meetings, Slack messages, and other virtual interactions. When permission was given, we also recorded virtual meetings and transcribed them using either Otter.ai or Rev.

Our team met with the senior director's task force on Zoom weekly. To maintain a close connection with PANL's leadership, the research team also joined the senior management team (SMT)'s bimonthly meetings. Researchers also attended meetings specific to the selected case studies, meeting regularly with managerial sponsors for the case studies, attending public events and town halls (all held virtually), participating in standup meetings and check-ins, and attending regular internal meetings (e.g., the IT team's weekly sync). Finally, the researchers shared a Slack enterprise grid with PANL and were able to join channels to observe informal chatter.

One of the advantages of joining as participant observers while studying remote work is that adding researchers to the meetings was information rich and relatively unobtrusive --- we could simply log in without disrupting the normal course of the meetings, and we had just as much information as any other remote employee. On the other hand, the digital ethnography made it difficult to get a sense of the culture at PANL, particularly its pre-pandemic atmosphere. While local team members made occasional visits to campus, and our interview questions asked employees to describe how the organization had changed since the remote shift, these efforts could not fully recreate the organizational culture, much of which had been tied to the physical site. Our digital ethnography was also focused only on specific teams and sub-units of PANL that were deemed amenable to some amount of hybrid work; we did not adequately observe or interact with, for instance, technicians and environmental safety specialists whose work is necessarily onsite. We recognize these as limitations associated with our method.

\subsection{Interview Design and Procedure}
The research team conducted 60 one-hour interviews with employees of PANL. The interviews were semi-structured, following previous interview study designs~\cite{kelly2017demanding}. Interview questions were iteratively designed while working closely with PANL stakeholders, with questions drawn from conversations around PANL's work design needs. Our protocol is provided in Appendix~\ref{appendix-interview-protocol}.

The interviews were conducted in four sets: (1) An initial set of interviews of 26 employees, selected by the senior director's task force as representative of the lab and its remote work experiences (labeled \textit{TF}); (2) 14 interviews with each member of the senior management team (labeled \textit{SMT}); (3) 14 interviews with members and key stakeholders of the IT group (labeled \textit{IT}); and (4) 6 interviews with members and key stakeholders of the Supply Chain group (labeled \textit{SC}). Note that the IT and SC interviews include both members of IT/SC and members of stakeholder teams that collaborate with IT/SC; an interviewee labeled ``IT \#'' may, in fact, be a stakeholder rather than an IT employee, and we will make the distinction clear in context. 
The initial participants for TF, IT, and SC were recruited through the task force; subsequent participants were recruited by snowball sampling from the original participants.

\subsection{Survey Design and Procedure}
The Chief Information Officer (CIO) of PANL provided a list of initial questions about the use of collaborative tools at PANL. A researcher on the team formalized the survey using techniques drawn from previous literature, and the CIO approved the final result (provided in Appendix~\ref{appendix-survey}).

Questions in Section 2 (the main section on tool use) of the survey were based on a previous survey-based study published in CSCW~\cite{zhang2020data}. The survey had asked workers to provide free-text responses about collaboration tools used in different data science activities. Our team adopted the same wording as this survey. Questions in Section 3 of the survey asked participants to fill out a technology overload scale. The scale was original, but drew from previous survey instruments on job stress~\cite{shukla2016development} and burnout~\cite{maslach1986maslach}.

The survey was distributed by PANL's internal communications team and received $198$ responses from full-time employees. Respondents had a mean tenure of $12.39$ years (median = $10$) and represented a diverse cross-section of the laboratory: 62\% worked in scientific groups and  38\% in support groups.

\subsection{Data Analysis}

\subsubsection{Ethnographic and Interview Data} Following the approaches of similar papers in the field~\cite{boehner2016data, lazar2018negotiating, gach2017control}, we fully transcribed the ethnographic and interview data, and collected them with field notes and emails. We then read and coded these materials using inductive grounded theory-based open coding. At least one coder reviewed all of the data, and themes were cross-checked and verified by a second.

After the initial inductive pass of coding, the first author deductively grouped the inductively-generated themes from the first step into the Distance Matters framework, organizing relevant data into tables~\cite{miles1994qualitative,cha2006values}. This second pass of deductively sorting data into an overarching framework aligns with previous work in CSCW~\cite{fiesler2016archive, tabor2017designing, chung2017personal}. These themes were then revised and verified by a second researcher in collaboration with the first.

\subsubsection{Survey Data}
Survey data was analyzed in R. An initial cleaning pass of the data consolidated similar or redundant free-text responses; for example, ``Microsoft Teams'' and ``Teams'' are considered the same response. The analyses were all exploratory (none were preregistered), but included analyzing the top frequently-reported preferred tools; measuring the spread of use across different tools; and understanding tool use given a particular sub-organization or occupational role.

\section{Findings}
Our findings proceed in two parts. We first present empirical evidence that strong intra-team remote collaboration per the themes of the Distance Matters framework may nevertheless lead to weak inter-team collaboration --- a \textit{Distance Matters paradox}. To explain the paradox in the second section, we expand the Distance Matters framework to account for the inter-team level of collaboration. We argue, using evidence from our field study, that the same socio-technical concepts from the original framework hold in a new form. These themes explain both PANL's successes and its challenges in adapting to remote work. In the Discussion section, we integrate the intra- and inter-team layers of the framework, highlighting how the two levels interact.

\subsection{The Distance Matters Paradox}
To provide evidence for the existence of a Distance Matters paradox, we begin with the story of the Central IT team, a team whose intra-team success belied its inter-team struggles.

\subsubsection{The Intra-Team Success Story}
IT comprised 80 members, divided into multiple sub-teams and specialized groups (see Figure~\ref{fig:it-org-chart}). The leads of all sub-teams and specialized groups met weekly with the CIO.

In general, the IT team members were very open to remote work. According to the CIO, approximately 60\% of IT's work could continue to be performed offsite post-pandemic. Many of its specialized groups, such as Cloud/SaaS Administration, IT Support, Communications, and Software Development, had an even higher estimation, with between 70-90\% of their work eligible for offsite completion. IT was also accustomed to communicating remotely; as Central IT manager (IT4) explained, ``a large portion of our [user base is] international anyway;'' the division even had senior members who were longtime remote workers (one interviewee, the lead of Unix Administration, moved out of state in 2016). These qualities allowed IT to lean into existing remote work infrastructure at the pandemic's onset.

As a result, IT adjusted well to working remotely. Within specialized groups, IT employees had loosely coupled work, balanced with a rhythm of check-ins and interaction. Each member took on designated roles and coordinated dependencies using shared software systems. According to the Network Engineering lead:

\begin{quote}
 ``Each of us kind of takes leadership on certain...aspects of our, of the things that we're working on. But I oversee that. And so part of that is project management, part of that is status updates, part of it is reporting to our management. And that reporting can take the form of email...we also use a system called Confluence to keep track of various projects and project status. It's kind of our internal documentation tool as well for much of our networking and network engineering.'' (IT2)
\end{quote}

While IT members noted some friction in translating in-person interactions to video conferencing, some noted that video calls eventually became ``a good thing. I can move from one meeting to another in seconds now, where if I had meetings in different buildings... I'd have to allocate certain amount of time.'' (IT2) Regular video meetings helped to ``maintain that camaraderie'' and included ``a little bit of joking around;'' team members felt very connected over shared technology, and were ``constantly talking'' (IT5). IT had been quick to adopt Slack for internal communication and troubleshooting, and employees found it very effective:
 
 \begin{quote}
 ``And Slack very quickly establishes itself as the chat platform for real time troubleshooting...whenever we have a major incident or event...they're just, you know, chatting back and forth and they're troubleshooting in real time. And, you know, that's really when it shines.'' (IT4)
\end{quote}

From the intra-team perspective then, IT employees appeared to be cohesive and productive. As the laboratory's experts on technology, and as the group with one of the lowest proportions of on-site work required, they were perhaps the best-equipped for transitioning to remote work. Regular meetings within sub-teams and specialized groups, and between the leads and the CIO, established common ground; jovial, informal interactions contributed to collaboration readiness; work was loosely coupled, yet well coordinated; and while the CIO had newly joined the lab a few months before the study, employees we interviewed found him very ``engaged'' (IT15, Local Technology Specialist), as well as ``open and communicative'' (IT4) with planning. The CIO had made such an impression while working almost entirely remotely --- a testament to IT's remote working capabilities.

\subsubsection{The Inter-Team Challenges}\label{inter-team-level}
At the inter-team level, however, the IT employees struggled to collaborate. Over the course of our study, IT encountered numerous difficulties engaging with scientific teams. Many science teams felt frustrated interfacing with IT on remote work-related issues, with some interviewees admitting to avoiding engagement with IT whenever possible: ``we don't even bother trying'' (IT11, Science User).

While tensions with IT certainly predated remote work, our data suggested that distance exacerbated teams' relationships with IT by replacing in-person exchanges with the IT Help Desk. A senior manager reminisced about being able to casually interact with support staff, allowing her to build an informal network. In the pre-pandemic days, this leader had a ``trusted guy in IT'' whom she would see around the office, but this interaction disappeared during COVID.

\begin{quote}
 ``People are siloed in their areas. You know, there are people who I would say hi to everyday, I don't even know [their names], but I know their faces.'' (SMT15)
\end{quote}

Another scientific manager similarly complained that IT support had declined, and had been replaced by a more one-size-fits-all approach.

\begin{quote}
 ``If somebody is going to work remotely... make sure that ... the IT support for them, from their keyboard, all the way to their, you know, to the work they're doing, is just as good as it would be on site.'' (TF18)
\end{quote}

At the lowest level of the hierarchy, a newly-hired summer intern (IT16) struggled to connect to the VPN; without an in-person contact to speak with, she wrestled with the IT Help Desk for days, and was assigned three separate incident numbers without resolution. The intern felt frustrated and disconnected; each time she was assigned a new IT employee, she had to ``re-explain'' herself.

In this way, distance drove a wedge between IT and its partners, making it more difficult for the two to communicate. Interviewees described how interactions with IT felt like speaking a foreign language. Scientists and scientific IT specialists described having to play ``a game of telephone back and forth'' (IT11, Science User), ``spend a lot of time explaining'' (IT10, Science User), and ``translate...information'' from the ``scientific'' language to the ``enterprise'' IT language (IT9, Local Technology Specialist). When the translation efforts grew too frustrating, scientists tended to take matters into their own hands: ``if other people have issues, sometime[s] they try to fix it themselves'' (IT10, Science User).

Challenges with communication also resulted in Central IT feeling disconnected from those they supported. One interviewee described struggling to support workflows that they did not understand:

\begin{quote}
 ``I can't be responsible for someone else's workflows that if I don't even know what it is, I can't even offer a solution.'' (IT1)
\end{quote}

A senior leader in Central IT predicted that the disconnectedness may also lead those outside of the IT organization to misunderstand the reality of its inner workings:

\begin{quote}
 ``I think that, generically speaking...the customers of our organization kind of treat IT as one big organization. ...And inside of PANL IT, we have these groups, and the groups have, you know, different management, and the groups do work together. ...now, from the outside, we're one big organization, but when somebody comes in with an issue or a problem or a service request...they may not realize there's ... all of these other teams that ... work together.'' (IT2)
\end{quote}

A previously regular annoyance for this leader had involved him being physically approached by scientists with questions that were outside his realm of expertise. He wished instead that more people would use official channels like the Help Desk ticketing system, which he described as ``a reasonable set of procedures in place to get the most critical services delivered in what I'll call a reasonable timeframe.''

Scientists did not seem to agree. They were frustrated by the prospect of now having to send their questions into the void, reported stories of slow and frustrating interactions, and preferred to have a designated point person instead:

\begin{quote}
 ``We don't really have the proper person that understands our systems and our labs that we can say, `Oh, yeah... I know who it is.' ... So there's just kind of this... void between us and them.'' (IT11, Science User)
\end{quote}

Although IT was very effective collaborating at the intra-team level, the inter-team level tells a different story: of teams that lacked common ground (and even a common language to communicate in); avoided collaboration when possible; felt uncomfortable with existing collaboration technology systems; and had misaligned expectations about the reasonableness of response times. That the same team could have such drastically different outcomes highlights the salience of the two levels of analysis. The inter-team level evidently poses a different set of socio-technical solutions than the intra-team level, and solving one guarantees nothing about solving the other.

\subsection{Extending the Distance Matters Framework}
The challenges that the IT team experienced at the inter-team level echoed the same themes from the original Distance Matters framework. This is no accident; in this section, we shall demonstrate why. We review the four themes of the original framework, as well as organizational managerial aspects (added later~\cite{olson2013working, bjorn2014does}). For each theme, we describe how the same underlying principles are transformed from the intra-team to the inter-team level.

\subsubsection{Common Ground}
The intra-team definition of common ground refers to the knowledge shared between participants. Very little of this definition needs to be adjusted for the inter-team level; the difference is that the participants now belong to different teams, rather than to the same one. Critically, inter-team common ground cannot be presumed from having intra-team common ground. The boundaries between teams, as well as their divergent backgrounds, culture, norms, and practices, may complicate the ability of the two teams to understand each other.

In IT's interactions with its scientific partners, for instance, the two teams often demonstrated a lack of common ground. IT had less familiarity with scientific jargon (hence the need for ``translation'') and approached problems with mental models that were incompatible with those of the scientists. An employee in a scientific organization described the frustration they felt brokering misunderstandings while working with Central IT to deprecate an older tool:

\begin{quote}
 ``Every alternative solution that was raised was inadequate to meet the needs of our groups or a lot of the groups that were operating at PANL. And...again, it just came back to another example of not understanding the use model of how systems are used in the experimental and development areas. So it's like, it's a continuous process.'' (IT11, Science User)
\end{quote}

In this way, IT struggled because its employees were generalists meant to serve broad enterprise functions, while its scientific partners were specialists with very precise needs. Lack of common ground made both parties feel that they were talking past each other.

On the other hand, we encountered a very different environment with the SC team. While, on the surface, the SC team was quite similar to the IT team, we observed the SC team thrive during the remote work period. In many ways, data from the SC team served as a counterpoint to data from the IT team: the SC team had chosen a very different inter-team approach, and thus created a fruitful contrasting example.

In particular, the SC team adopted a unique structure that enabled their members to achieve a high level of common ground. Rather than rely on a team of generalists to address customers' procurement issues, SC leveraged a model of embedded boundary-spanners. The embedded boundary-spanner (EB) would formally be an employee of the partnering team, and would ``sit in'' another sub-organization; however, their day-to-day role would be to interact with and pass on requests to the relevant individuals in SC. In effect, the EBs were ``bilingual,'' fluent both in the language of local groups and with that of SC. As one of the designers of the EB program explained, the goal is for an EB to ``speak a little bit of both [languages]'' (SC10).

The EBs we interviewed were proud of their role, and often had decades of experience working alongside the scientific teams in addition to experience with supply chain and procurement. The history of collaboration was significant; the original Distance Matters paper points out that ``people who have established a lot of common ground can communicate well even over impoverished media'' (p. 161). From this standpoint, the EBs had an advantage from the starting line: 

\begin{quote}
 ``I knew everybody, up, down, I knew the projects, the needs, I used to resource the computing. ...So, I know very intimately the details of how the organization works and its other directory. And a lot of people have told me, `Well, if I call the Help Desk, they don't know what I'm talking about.''' (SC3)
\end{quote}

EBs also had special technological privileges in the central system that tracked procurements. EBs wielded the ability to ``override most of the information that goes in [the system]'' (SC10), and could use their privileges, for example, to make context-specific corrections. EBs also had additional visibility about the status of procurements in the tracking system, which may have given them added value as a go-to resource, particularly in contrast to IT's generic ticket system.

As a result, the SC team was far more effective than the IT team in fulfilling its partnering teams' requests. During the pandemic, SC earned its highest performance ratings in seven years (93/100), achieving an ``Oustanding'' for the first time. Customers of the organization reported being able to trust the Supply Chain team:

\begin{quote}
 ``I feel like ...the people I work with are part of my team. So I go to somebody in procurement, they want to help me make happen what I need to happen.'' (SC1, Collaborator in the Environmental Protection Department)
\end{quote}

These findings highlight the importance of having common ground in facilitating inter-team collaboration. The EB is an example of a deliberate design choice that combines an organizational decision (embedded boundary-spanning) with a technological decision (giving special privileges in the central system) that facilitated improved common ground between teams.

\subsubsection{Collaboration Technology Readiness}\label{collabtechreadiness}
The intra-team notion of collaboration technology readiness (see Table~\ref{tab:dm-originalconcepts}) centers around the norms for using technology within a team, and the extent of its technological literacy. At the time that ``Distance Matters'' was written, some companies had yet to implement videoconferencing and messaging tools; part of the paper's argument was that organizations should adopt simpler technologies (e.g., email) before they adopt more advanced ones (e.g., video chat). Similarly, many organizations also lacked adequate IT support for the new software being introduced.

Today, however, the landscape has changed. Videoconferencing and group messaging are now ubiquitous in many organizations; even the most ``advanced'' features listed in~\citet{olson2000distance}'s original paper, ``hand-off collaboration'' (e.g., tracking changes) and ``simultaneous collaboration'' (e.g., screen sharing) are now standard, rather than novel. Today, far richer interactions are possible, including chatting both in real time and asynchronously (e.g., Slack, Microsoft Teams); using virtual whiteboards (e.g., Figma, Google Jamboard, and Miro); navigating shared virtual environments (e.g., Teamflow, Gather.town); managing projects (e.g., Trello, Jira, and GitHub project boards); and even creating sophisticated end-to-end workflows (e.g., Monday.com). Accordingly, some have declared victory over collaboration technology readiness at the team level~\cite{wolf2008does, leonardi2010connectivity, millernyt2021}.

The definition of collaboration technology readiness changes at the inter-team level. Being collaboration technology ready does not merely entail being fluent in technology --- this is now assumed --- instead, it requires being fluent in the \textit{same} technology (or at least inter-operable technology), having workflows that are visible and intelligible to collaborating teams, and having norms for data storage and knowledge sharing that cut across teams. The notion of technology support (mentioned in the original paper) continues to be relevant; the number of technologies used within an organization should not exceed the number of tools that IT can feasibly support.

At PANL, the sheer number of technologies in use signalled a lack of inter-team collaboration technology readiness. Despite surveying just 10\% of the lab's workforce, we found 9 distinct communication tools, 21 distinct tools for collaboration and project management, and 26 unique data storage locations. The survey findings aligned with interview data, in which informants cited a lack of technological norms, sparse knowledge management practices, and distributed collaboration systems. ``I think that's kind of a mess, to be honest with you,'' said one interviewee (IT6, Science User). Another stated, ``There's not a lot of standardization of anything'' (IT7, Science User). Throughout meetings, interactions, and interviews, IT showed multiple signs of being worn thin by supporting too many tools. At a cross-functional laboratory meeting, the CIO said:

\begin{quote}
 ``I mean, obviously supporting eight tools when you only need three, there's, you know, you're going to support those eight, a little bit less effectively than you would those three.''
\end{quote}

Even scientists were frustrated when working on multiple projects, because the data storage procedure for each project was sometimes determined on a project-by-project basis. When certain critical individuals left the institution, information about their data could be lost. One interviewee in a scientific group described having to call a colleague who had already left the laboratory to ask about their old data:

\begin{quote}
 ``Like somebody will quit, and then everybody's like, `Where's all the data?' You know ... we try to get people to put it all in one spot, but there's no like standard place where like when you quit, this is the procedure of where you put all of your data ... I mean, again, the guy who quit was like a good friend of mine. So luckily, I could call him afterward and be like, `what is this? Where do I find this?' and compile as much as he could. But yeah ... I can imagine if it wasn't somebody who, you know, left on good terms.'' (IT7, Science User)
\end{quote}

The large number of tools and lack of consistency in technology norms strained IT's ability to complete its job well. The team did not have the capacity to understand and support such a large number of tools and workflows. More generally, weak inter-team collaboration technology readiness made it harder for everyone --- even scientists --- to collaborate on projects across multiple teams, since those teams could have incompatible technology norms.

In this respect, SC had chosen a different approach, since a massive previous consolidation effort had merged its technological systems into a single tool, PeopleSoft. Thus, there was a greater ability to have a single source of truth, although collaboration technology use remained imperfect. Some employees used their own supplementary management systems, such as spreadsheets, and noted examples of alternative systems that they would have preferred (SC2); others complained that PeopleSoft was difficult to use and ``not intuitive'' (SC9, Environmental Safety Collaborator), and admitted that the system was not being used to its fullest capacity. To SC's credit, however, the team had invested heavily in its unified technology systems; its team website contained a ``Training'' page with digital tools and specialized checklists for each scientific team it supported (another signal of its commitment to understanding local scientists' needs). SC's focus on technology training, as well as its investment in EBs with special technological privileges, afforded it advantages over other teams that were spread thin over multiple tools.

Naturally, then, one might ask: why not always consolidate? At PANL, we observed several inter-team challenges that hindered collaboration technology readiness. As much as technological fragmentation was universally bemoaned, for example, teams at PANL nevertheless found consolidation nearly impossible, because each one was beholden to different external stakeholders. The Central IT team observed in a meeting:

\begin{quote}
 \textit{Person 1:} ``I've noticed that the groups at the lab that mainly interact...with [the] Department of Energy --- so, cyber, supply chain, finance --- they tend to use the DOE type suites, Microsoft Teams, SharePoint, OneDrive.'' \\
 
 \textit{Person 2:} ``...and the folks that interact mostly with the university on the research side and join[t] institutes and most of the academics, tend to use the tools that [the university] favors ...so it's Slack and Google Drive and Zoom.''
\end{quote}

Additionally, groups had customized tools to fit their needs. When asked why they would not be willing to change their tools for the sake of consolidation, one survey respondent wrote:

\begin{quote}
 ``The productivity loss would be massive; it's not just for me as an individual, but whole teams will be handicapped. Operational productivity hacks have been enabled in Slack to make us more efficient while remaining effective.''
\end{quote}

Furthermore, some roles were required to be compliant to specific legal standards, which limited viable technology options. Despite indicating that they were unhappy with their current tools, for instance, a survey respondent wrote that they were unwilling to switch because ``another tool will have to meet the same DOE monitoring and compliance requirements.''

Partially as a result of having different stakeholders and occupations, collaborators also had competing assumptions about technology. IT began every conversation with a discussion about requirements, assuming that each piece of hardware or software addressed a specific need. Meanwhile, local groups saw technology as part of an experimental process, and trial and error required taking ``a leap of faith'' (IT9, Local Technology Specialist). The differences in assumptions became a source of tension: Central IT would be frustrated, for instance, when other groups started a conversation by asking, ``Can we try this tool?'' rather than framing the discussion around requirements. As the CIO said in a meeting:

\begin{quote}
 ``What I didn't like about that conversation is you started with the tool, not the requirement, which in my world is an absolute no-no, you start with the what is the problem [you're] trying to solve. And then...figure out, okay, what tool is best.''
\end{quote}

In this way, adopting shared collaboration technology can be hampered both by hard requirements (e.g., legal compliance or stakeholder norms) and soft social norms (having different approaches and assumptions about technology).

Finally, technological literacy was relevant in our data, but played a small role. The IT employee who had been working remotely for years recalled his shock in observing colleagues who were less accustomed to remote work struggle to set up their workstations:

\begin{quote}
``I mean, a lot of people in Central IT very naturally do a lot of work from home. So I mean, we're very used to that. ... I did see a lot of tickets about just people not understanding how to just how to even do anything ... which was a little confusing to me, because I was like, you just do the same thing that you do at work. It's just, but I mean, I guess some people just never worked from home. So they maybe didn't even have a good computer at home. So they were just starting from scratch.'' (IT1)
\end{quote}

The collaboration technology themes we outline here each raise fundamental socio-technical questions that organizations should address if they attempt to scale up remote work. Questions like, \textit{Which technologies do we support?}, \textit{How much customization do we allow?}, \textit{Where do we store the data?}, and \textit{If, and when, do we allow exceptions to the company-wide technology standard?} should be asked as early as possible and communicated prior to creating a remote work policy. The organizations should have a sense for where they want to draw the line --- otherwise, teams will create their own norms or attempt to follow external norms, taking the control out of leaders' hands, and creating technological fragmentation that is more challenging to repair. 

PANL took steps to draw this line over the course of our study. After working with a member of our research team, PANL's IT group published explicit collaboration technology norms for the first time on its internal website. As of writing, the website has been live for approximately two months, and it is too soon to tell whether practices have sufficiently shifted. However, the website is an example of an artifact that deliberately balances intra-team customization with inter-team coordination. The aforementioned challenges had made it impossible to creating binding norms; instead, the IT group divided the tools into tiered categories: ``Recommended,'' ``General Support,'' ``Limited Adoption and Support,'' and ``End of Life'' (tools that would be decommissioned within two years). These categories enabled IT to slowly consolidate its tool portfolio by nudging groups towards recommended tools, while still allowing groups sufficient optionality.

\subsubsection{Coupling of Work}
At the intra-team level, coupling of work refers to the ``extent and kind of communication required by the work''~\cite{olson2000distance}, in which loosely-coupled work is sometimes (but not always~\cite{bjorn2014does}) more amenable to remote collaboration. 

At the inter-team level, however, coupling of work should be reframed as the temporal coordination required to connect the responsibilities of each team to the larger whole. In a sense, inter-team collaboration is already ``decoupled'' because each team is independently responsible for their portion of the project. The challenge is for teams to coordinate, identify dependencies, and refrain from blocking one another.

A key piece of the coordination involves shared expectations for timelines. Recall from Section~\ref{inter-team-level} that IT believed that it responded to Held Desk requests in ``a reasonable timeframe,'' while scientists complained of slow response times. It had turned out that IT found it ``reasonable'' to respond in a few business days, while scientists' experiments required 24/7 support.

Notably, the disconnect in perceptions of promptness had long predated the pandemic. It was because of the mismatch in temporal expectations that, years ago, science groups had established procedures to manage mission-critical computational processes internally. As one scientist recalled:

\begin{quote}
 ``And we tr[ied] to engage with office of the CIO at the time, but [we need to] provide that 24/7 support for the experiment... we ended up doing IT by ourselves.'' (IT10, Science User)
\end{quote}

Another piece of coordination involves planning appropriate times to interact. Members of scientific groups complained that IT would schedule maintenance at inconvenient hours. Scientists required very specific maintenance windows, which did not align with those of IT. Some groups had, over the years, even evolved to have alternative maintenance schedules:

\begin{quote}
 ``Maintenance windows are different at the enterprise level. So [for] enterprise...once a month at Tuesday, you have to patch your system from Microsoft. And that's the schedule. Well, when you have a user running, you can't just kick the user off at 10 o'clock at night, because they're running 24/7. So we have our own maintenance schedule as well. It's not to say that work doesn't get done, but it doesn't get done according to the enterprise schedule.'' (IT9, Local Technology Specialist)
\end{quote}

Both of these work-coupling issues predated the pandemic, so they are certainly not unique to the remote work context. Rather, remote work exacerbated their impact --- making it more difficult for IT and science groups to diagnose issues when systems failed. When our researcher initially partnered with the CIO, for instance, Central IT had hoped to understand why so many scientists were complaining about remote connectivity issues. The researcher soon discovered that technology work was so decoupled at PANL that it was difficult to identify the point of failure. 

In one interview, a science user (IT10) complained that remote network access was incredibly slow (``if we add the VPN connection, which is like the bottleneck before accessing all of this, then you can imagine how slow it can become sometimes''). When interviewed, Central IT's Networking lead (IT2) responded that the scientists were using the network inappropriately (``[our] VPN service is...not for research and science purposes''). In further interviews, the researcher discovered that individual groups had methods of remote access that varied wildly, with each managed by local technologists. And by the time Central IT established a coordinated response, local groups had created a workaround and moved on with research. Central IT had done too little, too late.

Coupling of work at the inter-team level is thus closely tied to common ground and inter-team communication channels. Because Central IT's timeline was so decoupled from that of the science groups, it was unable to establish an effective response for the network issue. The problem was exacerbated by the lack of communication about each team's unique practices and the distribution of technical responsibilities between Central IT and local technology specialists. 

When comparing IT's approach with SC's EB model, the contrast becomes even more stark. Unlike in IT, where the Help Desk would respond after a few days, SC's embedded boundary-spanners consolidated all supply chain-related information and responded instantaneously. The EBs were in lock-step with the groups they supported.

\begin{quote}
    ``We're losing what I used to call the lag, no, the time between getting the ticket and reaching out to the user...I believe the responses are better directly with a user and he can express himself. You can make them feel like we're here for you; he can see you. You can see him, even though you're not next to him, but it's feeling like, `yeah, we're here to work, we're supporting you. We're working with you.' '' (SC3)
\end{quote}

Thus, it is important for teams to build common work cycles, with both the technology (e.g., software updates and maintenance) and the organization (e.g., the rhythm of collaboration) centered around a shared understanding. As SC3 explains, a critical aspect is creating a sense of, ``We're working with you'' --- especially when the teams are working out of each other's sight.

\subsubsection{Organizational Managerial Aspects}
At the intra-team level, organizational managerial aspects often refer to management and planning decisions, such as goal alignment and planning. At the inter-team level, a similar principle applies: as our examples from previous sections have shown, systems of teams must coordinate their activities in order to succeed at working remotely, and planning is critical to this endeavor. In their review of multi-team systems,~\citet{zaccaro2020multiteam} find that the literature supports the creation of coordinating bodies with membership spanning the constituent teams, and that the bodies help coordinate systems with high \textit{component team diversity} (differences in geography, function, culture, or discipline). Boundary-spanning structures can especially serve to make coordinated decisions about collaboration technology, as they are boundary objects that can unify dispersed IT planning~\cite{desanctis1994coordination}. 

Our data suggests that these boundary-spanning structures may be effective. PANL's new CIO organized several such groups within a few months of his arrival. Eager to demonstrate his ability to listen to scientists, he created the PANL IT Council. The council contained managerial representation from both Central IT and each of the scientific groups, and met bimonthly to discuss technological issues relevant to the entire organization. Similarly, the CIO launched an effort to unify identity management at PANL. At the time of the study, identity management at PANL was partially managed by Central IT, and partially managed by scientific groups, leading to a fragmented login experience with multiple layers of authentication. The first step in the unification effort was to create a 30-person working group that spanned all of PANL, which was accomplished to some fanfare. The CIO reported on its launch during a sync with IT leaders:

\begin{quote}
 ``So, actually ... the biggest thing is probably the big splash we did with the IAM working group, where we had kind of on the order of 30 people from, like, intentionally the broadest cross section of PANL ... that I've ever ... met so far. So, great penetration, a lot of engagement.''
\end{quote}

By June, the working group had produced a white paper, although as of the writing of this paper, the identity management project had not yet reached completion.

At a high level, the organizational theme for inter-teams requires coordinating leadership between multiple groups, and establishing consistent policies and practices to guide shared activities. These policies and practices serve as the guardrails of collaboration in any distributed organization. We caution, however, that guardrails do not replace efforts to connect teams on the ground. The CIO admitted, for example, that while IT and science had become more engaged at the leadership level, the Council was playing ``a big giant game of telephone.'' ``It’s four levels of disconnect by the time the person on the ground gets the communications ... a lot gets lost in translation.'' Therefore, the organizational theme must be considered in tandem with the others: coordinating leadership does not absolve the need to consider the common ground, collaboration readiness, collaboration technology readiness, and coupling of work for the different teams.

\subsubsection{Collaboration Readiness}
The intra-team definition of collaboration readiness refers to a willingness to share information that aligns with the incentive structure. As with common ground, this definition is also largely relevant at the inter-team level; however, the ``incentive structures'' will involve between-team, rather than between-member, constructs. Challenges may include misaligned goals~\cite{sherif1961intergroup, marks2005teamwork}, a lack of team identity~\cite{connaughton2012social,bos2010shared}, unequal status~\cite{sherif1961intergroup, metiuowningcode}, and invisible interdependencies~\cite{bick2017coordination}. Navigating these potential issues often requires levels of coordination beyond what is necessary to establish intra-team collaboration.

The theme of collaboration readiness was highly salient in our data from the start. When conducting our initial 26 interviews, 11 (42.31\%) reported having to work longer hours compared to before the pandemic, with eight (30.77\%) specifically highlighting challenges in inter-team collaboration, and eight finding it difficult to work with new team members.

Analogous to Olson and Olson's findings for within-team interactions~\cite{olson2003mitigating,olson2000distance}, interviewees characterized the shift to remote work as jarring, revealing structural gaps in inter-team collaboration that had previously been smoothed over by face-to-face interactions:

\begin{quote}
    ``I think [the most difficult thing] for me would be the lack of interaction, making it difficult for people to set up trust to set up new collaborations and to you know, stay aligned with with the mission.'' (TF6)
\end{quote}

Another participant described the lack of formal inter-team collaboration infrastructure, such as ``cross cutting meetings:''

\begin{quote}
    ``[We had] challenges to being able to maintain collaboration... across siloed organizations. Looking back, maybe we could have had, you know, put in regular meetings and things like this...maybe we could have had more effective cross cutting meetings and more open discussions and things that might have helped some of those processes earlier.'' (TF16)
\end{quote}

Similarly, those bringing on new employees found severe gaps in the formal documentation of PANL's internal processes:

\begin{quote}
    ``The checklists and things that they sent were confusing. Sort of outdated.'' (TF5)
\end{quote}

Because PANL had so heavily relied on person-to-person interaction, there had been little incentive to formally establish inter-team collaboration procedures. But remote work had laid its structural holes bare. Suddenly, asking questions to strangers was a much more frictionful process. Communication became much more deliberate; casual questions were left unsaid.

\begin{quote}
    ``Oh, I got some questions. I'll go over and talk to Joe about it. And you walk over...have a very fluid conversation. Whereas if you've got some questions [now], and you want to talk to somebody about it on email, you have to kind of gather your thoughts and put them down in a structured manner and effectively communicate it.'' (TF13)
\end{quote}

As a result, the frayed edges of relationships became more apparent. This was the case for IT-science interactions; when science users made attempts to contact IT for remote setup issues, their early interactions revealed the disconnect between their teams:

\begin{quote}
    ``Early on, I was complaining about my VPN connection. I mean, I literally had an IT guy say something back like, well, it works fine for me. And I'm sorry...am I making up my complaint?'' (IT6, Science User)
\end{quote}

The lack of understanding reflected broader issues of having misaligned goals and distinct organizational identities — challenges that had predated the pandemic, but became all the more apparent during the remote work period.

\begin{quote}
 ``I think it's just a lack of understanding from the start, you know, they get brought in thinking, Hey, we're gonna go work at the lab. And then they realize that they're in this more business environment. And then the scientific people are so disconnected from that business, that it's really challenging to bring the folks together. And I think that's one of the largest challenges.'' (IT9, Local Technology Specialist)
\end{quote}

As a business rather than scientific function, IT was also relegated to a lower status, and was seen as a less credible outsider when involved in the science teams' affairs. Its technical contributions in projects were often met with resistance, and Central IT worried that scientists were increasingly seeing their relationship as transactional rather than collaborative. One Central IT employee summarized the relationship by saying, ``Ultimately, this lab is run by scientists. It's not run by [IT]'' (IT4). Similarly, the CIO stated,

\begin{quote}
 ``There's, you know, [a mentality of] `IT doesn't know what they're talking about so we don't trust them.' And so... Central IT has very little agency or power or influence to enact any sort of prescriptive change or anything like that.''
\end{quote}

Thus, our data suggest that working remotely worsened strains between IT and its scientific partners, deepening existing divides across discipline and status. This finding is in line with previous research that remote collaboration can lead to an ``out of sight, out of mind'' effect that devalues the contributions of the lower status teams~\cite{metiuowningcode}. The story of IT also fits into the PANL-wide narrative of experiencing greater collaboration challenges across silos: distance attenuated the places where existing collaboration readiness was weakest.

\subsection{Summary of Findings}
Our results highlight key CSCW challenges in the design of distributed work, such as building inter-team, rather than only intra-team, identity; walking a fine line between technological unification and customization; facilitating the organizational and temporal coordination of work across different teams; and establishing a common language between teams with little shared context.

In the next section, we reflect on the tension between the intra- and inter-team levels; not only do these levels pose distinct design challenges that distributed organizations must separately address, but the solutions to these challenges also fundamentally trade off with one another. We integrate this discussion of the theory with clear practical implications for the ongoing shift towards hybrid and remote work.
 
\section{Discussion}
Over the next several decades, we anticipate that remote collaborations will grow increasingly complex. As global organizations face the prospect of permanent remote work, this paper extends a classic theory to account for the key challenges and design decisions for making such work sustainable and productive. We also raise the problem of a fundamental tension between intra- and inter-team collaboration; this tradeoff implies that there is no one-size-fits-all solution to remote work. Rather, organizations face choices. How should teams share knowledge with one another? Who gets to cross team boundaries, and when? To what extent should teams get autonomy, and to what extent must they follow organizational guidance? What are the exceptions --- for example, legal requirements for teams dealing with sensitive data? Each of these questions represents a potential tradeoff between the intra- and the inter-team levels.

\subsection{Is it Too Much to Ask for Both?}

\begin{table}[tb]
{\small
\begin{tabular}{@{} p{0.2\textwidth} | p{0.25\textwidth} | p{0.25\textwidth} | p{0.25\textwidth} @{}}
\toprule
\textbf{Concept} & \textbf{Intra-Team} & \textbf{Inter-Team} & \textbf{Tension}\\
\hline
 Common Ground & ``Knowledge that the participants have in common, and they are aware that they have it in common'' (p. 157). & Having shared knowledge and the awareness of that knowledge across team boundaries; teams understand each other's thought worlds and speak a similar ``language.'' & Too much common ground would lead to losses in individual teams' unique expertise, but being too specialized creates ``translation'' issues. \\
 \hline
 Collaboration Readiness & ``A willingness to share…[that] aligns with the incentive structure'' (p. 164). & Being willing to share information and work on joint deliverables. The teams have shared goals, mutual trust, and a shared team identity. Status differences do not impede productivity. & Too-strong global identity impedes a local sense of belonging, but too-strong local identity causes inter-team conflict. \\
 \hline
 Collaboration Technology Readiness & The ``habits and infrastructure'' around technology, including ``alignment of technology support, existing patterns of everyday usage, and the requirements for a new technology'' (p. 164-5). & Being fluent in shared technology, with visible and intelligible workflows and with norms for both data storage and knowledge sharing. & Hyper specialized tools lead to poor coordination, but generalized tools reduce individual teams' productivity. \\
 \hline
 Coupling of Work & ``The extent and kind of communication required by the work'' (p. 162). & Coordinating the teams' responsibilities and dependencies in a shared rhythm or timeline. & Aligning all work to the same schedule would create local inefficiencies, but failing to set scheduling expectations blocks local progress. \\
 \hline
 Organizational Managerial Aspects & ``What the manager can do to ensure that the collaboration is successful,'' including aligning goals, designing reward structures, planning, and communicating decisions (p. 49~\cite{olson2013working}). & Establishing boundary-spanning structures with membership spanning the constituent teams. The structures help to make coordinated decisions in areas related to the collaboration. & A boundary-spanning organization can be too strong with top-down policy, hurting the autonomy of individual teams; one that is too weak can fail to surface and align important dependencies, creating policy conflict. \\
 \bottomrule
\end{tabular}
}
 \caption{A summary of the fully-expanded Distance Matters framework. We present the intra-team level (the original framework) in conjunction with the inter-team level (our proposed expansion) and the tension between them.}
 \label{tab:interteam}
\end{table}

To start, one may rightly ask a practical question: is it possible to satisfy both layers of the Distance Matters framework at once? Thus far, our extension to the framework has been presented as a separate, but related, layer to the original. One could imagine the designer of an organization simply considering each set of themes in turn --- first satisfying the Distance Matters criteria within teams, and then satisfying the criteria between teams.

Reality, unfortunately, is more complicated: The intra- and inter-team layers of the framework are not sequential, but rather in tension --- making it difficult (though not impossible) to simultaneously satisfy conditions at both levels.
Table~\ref{tab:interteam} summarizes the fully expanded Distance Matters framework, and the between-level tensions inherent in each theme.
Consider the example of collaboration technology readiness. Teams satisfying the intra-team level \textit{too} well would create an incredibly effective local environment, but it would be too specialized, thereby impeding coordination efforts. Similarly, if the entire organization adopts the exact same set of tools (which would make IT very pleased!), the tools would be so general that they would ignore individual teams' needs. As the survey respondent pointed out, the productivity loss from eliminating specialized hacks and customized tooling would be massive.

The teams we observed found themselves with two choices: to compromise --- and adopt less-than-ideal tools for the sake of global coordination --- or to introduce redundancy --- and adopt multiple similar tools in order to accommodate everyone. Most chose the latter, resulting in the archipelago of collaboration technology that we observed at PANL. As a consequence, individuals who worked with multiple teams complained of an overload of similar technologies.

A similar tension exists across all five themes. An inter-team collaboration with too much common ground would lose individual teams' unique expertise, but being too specialized creates ``translation'' issues. Collaborations with a too-strong global identity will struggle to create a local sense of belonging, but too-strong local identity causes inter-team conflict (as noted by~\citet{dechurch2013innovation}). Collaborations that align all work to the same schedule would create local inefficiencies, but failing to set scheduling expectations blocks local progress. And finally, a boundary-spanning organization that is too strong can take autonomy away from constituent teams, but one that is too weak can fail to surface and align important dependencies.

\begin{figure}
 \includegraphics[width = 0.6\textwidth]{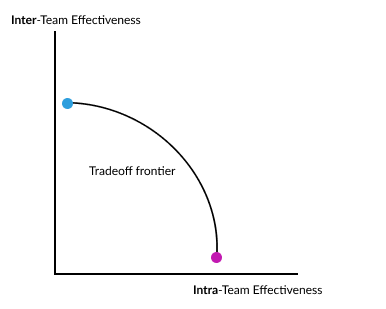}
 \caption{Diagram of a tradeoff frontier between intra-team effectiveness and inter-team effectiveness. When closer to the blue dot, the inter-team layer is more effective, at the cost of intra-team effectiveness; closer to the magenta dot, the opposite is true.}
 \label{fig:intra-inter-tradeoff}
\end{figure}

Figure~\ref{fig:intra-inter-tradeoff} summarizes the relationship between the intra-team layer (on the X-axis) and the inter-team layer (on the Y-axis). The tradeoff frontier illustrates that, given organizations' finite resources, it is easy to focus too much on one at the expense of the other; the design challenge of remote organizations is to determine where on the frontier they wish to position themselves. The challenge for CSCW, in turn, is to determine the shape of the frontier. As technology and social norms evolve, will the tradeoff become less prominent? What sorts of technologies could mitigate this tradeoff, if any?

What is clear from the frontier, too, is that, although intra-team and inter-team needs are in tension, they are not necessarily a zero-sum game. The SC team stood out as a team that managed to do both, and thrived while collaborating remotely (as evidenced by its high scorecard ratings both before and during the pandemic). The team is an example of one that shifted its frontier outward, increasing the size of the ``pie.'' 

Like every other team, the SC team still faced tradeoffs --- it had simply taken strong stances at each decision point. SC had invested, for example, in a unified technology system on PeopleSoft, even though several SC members and collaborators did not feel fully proficient or comfortable with the software. Instead of catering to local preferences, however, SC promoted an approach of making its one system fit all --- using customized training to adjust to local needs.

Additionally, the SC team established practices that increased the team's overall effectiveness, in both inter- and intra-team interactions. SC maintained a meticulous set of scorecards for customer success, which were based purely on inter-team performance (ratings from customer surveys). We conjecture that these scorecards created incentives for establishing strong inter-team collaboration structures, including the EB system; the use of metrics likely also created a culture of transparency and drive for improvement in the team's daily work. 

The heavy investment in training and transparency may have contributed to an overall increase in the team's effectiveness. However, future work should explore these possibilities further, as the specific implications of SC's metrics and other practices were not the primary focus of this research.

\subsection{Implications for the Remote Work Shift}
As with the original Distance Matters framework, we believe that the inter-team criteria we lay out for each concept are far more straightforward to achieve in an in-person context --- for instance, it is easier to sustain inter-team common ground when one sees the other team in the same shared office, than when that team is hidden behind technology. Therefore, as organizational leaders weigh whether, and how, to approach remote work, they should consider whether the organization is committed to explicitly creating the socio-technical conditions that enable remote work to succeed. As we have noted throughout this paper, this investment involves both an organizational strategy (for example, creating boundary-spanning roles and structures) and a technology strategy (for example, establishing shared collaboration tools with the desired settings and permissions).

Furthermore, organizations should not be satisfied with merely observing that their teams have produced good quality results during pandemic-era remote work. As Judith and Gary Olson noted in their original work~\cite{olson2000distance,olson2003mitigating}, distributed teams can often produce work that is of the same quality as co-located teams, but endure a far more painful process --- and require far more explicit management --- to achieve the equivalent result. High-quality results can obscure significant friction underneath the surface; at PANL, for instance, the laboratory continued to execute its scientific mission, despite significant challenges in collaboration technology readiness at the inter-team level. Over the course of our ethnography, we were consistently impressed to see people rising to meet challenges; however, tenacity cannot be mistaken for long-term sustainability. One of our interviewees early in the study described herself as being in ``survival'' mode, working overtime hours to meet important deadlines. ``Don't tell us that you want us to be more productive and more efficient,'' she said, ``because I'll just lay down and die'' (TF4). 

TF4 was not alone. Recall, too, that over 40\% of our initial 26 interviewees discussed working longer hours after transitioning to remote work. While some of these hours may be attributed to extenuating circumstances due to the pandemic, many of our interviewees were finding themselves forging cross-organizational connections anew when face-to-face interactions disappeared.

For remote organizations to have longevity, then, leaders must address the underlying tensions for common ground, collaboration readiness, collaboration technology readiness, coupling of work, and organizational managerial aspects. If PANL were to stay permanently remote, for instance, it would require a strategy that builds in ways for teams --- particularly those, like IT and science, with wide status differences --- to connect with one another and to build common ground. Otherwise, cracks in the relationship would widen into a chasm over time, and distance may further obscure communication challenges.

Organizational leaders should also not be satisfied if they observe that some individual teams have thrived with remote work. Teams do not exist in a vacuum; achieving effective work at the intra-team level does not imply an effective inter-team level. Thus, an important implication for this paper is that CSCW must not stop at designing for \textit{teams}' success --- it must ensure the success of \textit{multi-team systems}, and of the organization as a whole.

Here, the Supply Chain team (SC) serves as a noteworthy example. Our data show that the team thrived in remote work because it had a strong investment in inter-team relationships. Its internal website provided training for each sub-organization's needs, and the embedded boundary-spanners (EBs) deliberately ensured common ground and collaboration readiness between SC's stakeholders. SC's success during its fully remote period demonstrates that, when the socio-technical conditions are right, remote work can be highly effective --- perhaps even more effective than working in-person, given their stellar performance ratings.

Our results should therefore guide organizations to be thoughtful about how they approach remote work, and that the IT and SC case studies create guideposts for the types of decisions that leaders may encounter. We expect each concept in the framework (the rows of Table~\ref{tab:interteam}) to pose a key question for organizational leaders. For instance, in reading the first row of the table, a leader could ask, \textit{how might I ensure that knowledge is easily shared across team boundaries, while still allowing local teams to exercise their expertise?} We hope that this paper becomes a scaffold for making better-informed decisions about remote work.

\subsection{Technology Design Implications}\label{section-tech-implications}
Our research also reveals that the field is sorely in need of technological solutions to inter-team problems. The vast majority of collaboration technology on the market today is optimized to help teams succeed; teams can track their progress via Scrum, brainstorm ideas on virtual whiteboards, and discuss their results in increasingly rich virtual environments. However, we found few tools designed to facilitate inter-team coordination. The recommended tools for multi-team systems we found tended to be team-based tools --- for example,~\citet{anania2017communication} references Flow, Slack, and Teams, most of which were intended to be used for team-based project management. More generally, Scrum and other Agile methods were designed around small functional teams, and scale poorly in an inter-team context; in the organizations studied by~\citet{paasivaara2012inter}, attempts to adapt scrum for multiple teams (e.g., ``scrum of scrum'') fell flat. Employees found the meetings to be a waste of time, and the weekly rhythm of the ``scrum of scrums'' was too slow to meaningfully impact the rapid pace of day-to-day work (creating a Coupling of Work issue at the inter-team level). Similarly, a review of 42 case studies by~\citet{gustavsson2017assigned} found that the role of coordinating inter-team work varied widely between organizations, suggesting that designing for inter-team collaboration remains ripe for new technological solutions.

Current tools supporting intra-team collaboration do not scale well to the inter-team level. A Slack channel can be a lively place to discuss ideas with colleagues; a workspace with dozens of such channels, however, would be nearly impossible to keep up with. We anticipate possible tool designs that experiment with intelligent nudges~\cite{thaler2021nudge}, such as notification summaries, chat digests, or automated detection of related work --- nudging someone to say, ``hey, what you're working on might benefit from a conversation with a team down the hall!'' Technological scaffolds could also serve to align teams with very different temporal rhythms. In the same way that Gmail will nudge a user with, ``Sent 3 days ago. Follow up?'' a collaborative scaffold might say, ``The physics team works on a three-day cycle. If you don't follow up today, your suggestions may be lost.''

In this vein, a previous line of work has investigated tools that connect similar individuals in an organization using their digital traces, such as~\citet{carter2004building}'s Hebb System and~\citet{tang2007exploring}'s Consolidarity. Each of these systems uses digital traces of employees (emails in the former, all files in the latter) to identify individuals similar to oneself, who may be valuable connections. A possible extension of this work could be to use an algorithm to facilitate the employee onboarding process, and automatically connecting a new hire with relevant colleagues. All-remote employees could also be periodically paired with colleagues in adjacent or related areas (similar to the Donut Slack application, which sets up serendipitous coffee chats), and thus reduce the risk that distance would leave teams siloed.

Another area of exploration could be to design meta-tools that connect otherwise incompatible computational systems. As Section~\ref{collabtechreadiness} emphasizes, unified collaboration technology is often instrumental to effective inter-team collaboration, yet logistically infeasible for a variety of reasons. Is a ``universal translator'' across tools possible? Future tools may build on the work of~\citet{karger2005haystack}, who developed ``Haystack,'' a project enabling end users to flexibly view and retrieve data objects of different types. We imagine using similar methods of converting data from one system into a searchable format for another, thus reducing the friction when a Google Suite user collaborates with a Microsoft user. An example of an automation tool in this vein is Zapier~\cite{zapier}, which can automate actions on Slack from Microsoft Teams, and vice versa, to create a rough sense of interoperability. Perhaps more broadly, what does a virtual ``common room'' or gathering space look like, given teams fragmented across different technologies?

Ultimately, there is no technological silver bullet for bridging the inter-team gaps that distance creates. Any technological solution for one level of collaboration will grapple with potential trade-offs at another level of collaboration. However, the area is ripe for new ideas. With these suggestions, we hope to inspire technologists to build solutions that target inter-team collaboration issues, just as tools emerging in the last decade have addressed intra-team issues.

\subsection{Organizational Design Implications}

\subsubsection{Managing Teams of Teams}
At the highest level, our work underscores the additional demands of distance work introduced at the inter-team level. In managing teams of teams, organizations should establish regular mechanisms of engaging shareholders, formalizing shared goals, and coordinating activities. Just as intra-team collaboration requires planning, so inter-team collaboration must also have a clear plan and well-aligned rhythms. When the accountability structures support, rather than obscure, the problems being solved, the collaboration truly shines. 

\subsubsection{Increased Importance of Collaboration Technology Readiness}
Furthermore, our research finds that collaboration technology readiness is crucial. Unlike at the intra-team level, where collaboration technology readiness has recently become much less of a concern, different teams will have different technology stakeholders, occupational norms, and assumptions. The teams may use entirely different platforms, or use the same platforms in different ways. Technological literacy can also vary widely. There may be cases when it is impossible to unify distinct streams of technology --- when the constraints of different organizations are spread so far apart as to be impossible to consolidate on a single platform. In this case, organizations will need to facilitate multiple platform options without allowing them to become siloed; as we suggested earlier, cross-platform translation tools could be a fruitful area of exploration.

\subsubsection{Creating Socio-Technical Roles}
Finally, as the contrast in common ground between the IT and SC cases illustrates, different organizational designs can lead to drastically different outcomes. The embedded boundary-spanners (EBs) were far more effective at brokering relationships between SC and science groups because they were embedded in scientific teams, while IT employees were seen as outsiders by the scientists. EBs also had the advantage of being ``technologically embedded'' with privileges in the software itself. They are an example of a socio-technical role: a position that leverages both technological and social characteristics to achieve the desired working outcomes. We anticipate that other socio-technical roles may arise in the coming years to address the changing status of the office. One example could be an ``in-person surrogate'' who performs on-site actions on behalf of a fully remote employee. In meetings at PANL, we observed some managers being open to allowing fully remote workers, so long as other employees ``cover'' the on-site portion of the work. One senior manager described this perspective in her interview:

\begin{quote}
 ``I need at least one of them on site at all times. So whether they start cross training each other, whether they work out some kind of, you know, that type of schedule, they need that kind of coverage.'' (SMT13)
\end{quote}

At PANL, several scientific groups also began investing in augmented reality (AR) headsets, enabling an on-site individual to relay their live perspective while communicating with an off-site partner. This technology could help facilitate the creation of roles like an in-person surrogate.

\subsection{Future Directions}
We hope that, by articulating a new layer of the Distance Matters framework, this paper inspires future research to further investigate inter-team remote collaboration and the tension between the intra- and inter-team levels. Indeed, the challenge of how to balance the team's interests with those of the organization sparks numerous follow-up questions. What forces drive the intra-team and inter-team levels apart, and what forces can make them more closely aligned? How might future collaboration technology help ameliorate the tension? We imagine that many of these studies will be behavioral --- investigating and quantifying the tension, and empirically mapping the relationship that we roughly sketch out in Figure~\ref{fig:intra-inter-tradeoff} --- and others should be technological contributions, some of which we have discussed in Section~\ref{section-tech-implications}.

There should also be further explorations of the translation between intra- and inter-team constructs. We note, for instance, that common ground and collaboration readiness are defined in almost the same way at both levels. The only difference is the unit of analysis; at the intra-team level, it is the individual member, and at the inter-team level, it is a team. Collaboration technology readiness, on the other hand, needs to be conceptualized differently: rather than focusing on core competencies in using tools, the inter-team level focuses on systems of technology and shared workflows. It would be worthwhile to understand how the translation between levels shapes how we understand the constructs themselves. For instance, does the generalizability of the ``common ground'' concept make it more important? Are some constructs more important than others at a given level? We have previously discussed the example of collaboration technology readiness (which is more salient at the inter-team level than at the intra-team level), but more analysis is required for the other themes.

Additionally, it will be interesting to ask whether there are new constructs, outside the five described in ``Distance Matters,'' that are relevant for the inter-team level, or whether any of the constructs should be removed from consideration. In this paper, we tried to remain true to the original framework. Future studies should explore whether, and how, the framework should be adjusted.

Our study also raises a large number of questions both for organizations transitioning to remote work and for those that have already done so. Each of these questions will become increasingly important as more organizations make this transition. For instance, how does transitioning from remote work differ from starting as remote from scratch? Is it ever possible for an in-person firm to achieve the same level of fluidity as one that was ``remote first'' from the outset? Organizations may even explore a hybrid approach. We note in our study, for instance, that the successful SC team had a substantial history of collaboration, and that this shared history --- in addition to its socio-technical design decisions --- was instrumental to its inter-team relationships. Thus, a potential avenue for future work is investigating how mixing remote and in-person work impacts organizations' abilities to achieve the conditions we outline in the expanded Distance Matters framework. Continuing~\citet{olson2003mitigating}'s decades-long line of work, organizations can experiment with in-person retreats (already a staple at location-independent organizations~\cite{rhymer2020location}); as-needed co-working spaces; and hybrid schedules, in which employees are expected to be on-site for face-to-face activities during designated portions of the week.

Finally, further research should explore how the theory we extend holds in different organizational contexts, as well as longitudinally. How do the size, dispersion, geographic location, and demographic diversity of an organization play a role in its ability to achieve effective remote work conditions? How do patterns of communication and collaboration change over time, as an organization transitions to remote? How sustainable is a remote organization, and in what ways should the organizational and technological design evolve to make the changes last over decades? Each of these questions poses a potential avenue for future investigation.

\subsection{Limitations}
Overall, we are optimistic that the collaboration challenges we found at PANL represent those experienced by many inter-team collaborations. Our work builds upon decades of research on ``translating'' across organizational boundaries, interdependence between teams, and the dynamics of managing a group's boundaries. Thus, while we believe that the characteristics of team interaction we observe are not unique to PANL or only to laboratory settings, we also recognize that additional work will be required to prove the robustness and identify the limits of our theory in other settings.

An important limitation of this work is that the data was collected from a single organization during a period of transition due to exogenous shock (COVID-19). Although the timing and setting afforded us a detailed case study of this organization's adaptation to remote work, this singular focus is also a primary drawback of this research.

We recognize, for instance, that our field site was a large, English-speaking, decades-old scientific laboratory in a relatively affluent area in the United States. Organizations with different characteristics and different demographics, and in different contexts, may have different experiences when making a transition to remote work. We highlight in particular that, in studying remote work, the phenomenon is \textit{global}: we believe that this theory would benefit from evidence of international distributed organizations. We therefore strongly urge future researchers to increase the generalizability of our findings.

\section{Conclusion}
The Distance Matters framework has been influential in the CSCW literature for over two decades, and for good reason: in extending the framework to account for the inter-team collaboration, we ultimately find that the original five categories apply quite well to scaling distributed teams to distributed organizations. Our work demonstrates, however, that looking only at the team level can leave much to be desired at the inter-team level. In fact, efforts to optimize teams for remote work can undermine the effectiveness for organizations as a whole --- opening numerous future questions for organizations transitioning to remote work. Therefore, the expanded Distance Matters framework represents a rich new area for further study.
\begin{acks}
The authors sincerely thank our field site partners, without whose help, support, and close collaboration this research would not have been possible. Thank you for allowing us a window into your organization, and giving us a chance to participate in the process of navigating remote work. We thank all of the anonymous participants who generously shared their time in interviews with us. We also thank our anonymous reviewers for their incredibly helpful feedback and thoughtful commentary. Finally, we are grateful to the Stanford Symbolic Systems Program, particularly to Hyowon Gweon and Todd Davies, who co-led the Master's Research Seminar. We also thank the Stanford Management Science and Engineering Center for Work, Technology and Organization, as well as Wendy Arroyo Pazmino for her excellent research assistance. Finally, we thank Joseph Seering for providing extremely useful feedback on a draft.
\end{acks}

\bibliographystyle{ACM-Reference-Format}
\bibliography{bibliography}

\appendix

\section{Sample Interview Protocol}\label{appendix-interview-protocol}
Below is a general interview protocol that we used for interviews. This protocol was typically refined with additional role-specific questions.

\paragraph{Logistics.} 
Introduce the role of the task force and get permission to record.

\paragraph{Background.} 

First, I want to understand your main work. 
\begin{enumerate}
 \item What is the most important work that you are accountable for in your job? (Probes: timing, measures, evaluators of work)
 \item Who do you interact with the most for your work? Tell me a little bit about those interactions? (Probes: whether interactions with managers, mentors are important)
\end{enumerate}

For the next 4 questions, I want to focus on learning about the period of time right after PANL went to all remote. These questions are intended to help PANL understand how to better support you and your team right now. Then, I will ask about your ideas for future plans.

\begin{enumerate}
 \item Since the pandemic began, do you have a sense of what your hardest [unit of work mentioned above - project, etc] was in the context of your work at PANL? Can you tell me about it? (to help with the story, perhaps probe for: Where were you, when did it start, what happened? Who did you interact with and how? Why was it good?)
 \item In contrast, have there been any [projects] that have gone particularly well? Can you tell me about it? (to help with the story, perhaps probe for: Where were you, when did it start, what happened? Who did you interact with and how? Why was it good?)
 \item What new technologies has your team started using? How have those worked? (Can you give me an example of it working well? Can you give me an example of it working poorly?)
 \item What new routines or processes has your team started using? How have those worked? (Can you give me an example of it working well? Can you give me an example of it working poorly?)
\end{enumerate}

\paragraph{Needfinding.}
\begin{enumerate}
 \item I’d love to hear how you are structuring your work day, especially any frictions or challenges that are coming up. (Probe for interactions, frictions, failures)
 \item And how are you structuring your work week? 
(Probe for interactions, frictions, failures)
 \item Can you think of a time recently when you needed information about what it takes to do your job? How did you get that information?
 \item What is the biggest unmet need your team has right now?
\end{enumerate}

\paragraph{Comparisons Pre/Post.}

\begin{enumerate}
 \item How has the pandemic changed your most important work and accountabilities? Probe for: Have new responsibilities been added?
 \item Have prior responsibilities gone away? Or been put on hold?
 \item How are you held accountable for the new work?
 \item How has the move to remote changed your daily allotment of time?
 \item How has the move to remote changed your weekly allotment of time?
 \item I’d also love to hear about how the more social side of work has changed. Can you tell me about a time in the last three months where you felt particularly connected with your colleagues, outside of a work interaction? 
Probe for understanding: How do they socialize?
 \item Something that we have heard come up in meetings is the idea of “maintaining the culture” -- what does that phrase mean to you with regard to PANL?
 \item How has that culture changed with the move to remote? 
\end{enumerate}

\paragraph{Future of Work Policy.}
Now I’d like to transition to hearing your ideas about the future of PANL and the potential promises and challenges of a telework policy. We know that PANL is exploring local and global telework policies.

Let’s say for a moment that every manager got to define a telework policy for their team. (Have them describe who they think of as their team and manager if it hasn’t come up) What do you think an ideal telework policy might look like for your team? 

\begin{enumerate}
 \item What would be the main reason for that particular policy?
 \item What would be the main reason for that particular policy?
 \item What trade-offs do you think you’d be balancing there?
 \item What would be your main concern with that policy?
 \item How different is it from what you’re doing now?
 \item What resources or capabilities would you need to have in place for this to work that you don’t have now?
\end{enumerate}

OK, so that was a brainstorm about a local telework policy. Now I’d love to hear what you think might work for a global telework policy. Should every team have that same telework policy? Should every manager decide for their group? 

\begin{enumerate}
 \item What would go wrong if every group adopted that same policy?
 \item What would go wrong if every manager decided for their group?
 \item What are your main concerns about a PANL telework policy?
Do you have a sense for what you’d like to see?
 \item What resources or capabilities would PANL need to have in place for this to work that you don’t have now?
\end{enumerate}

For the final question, let’s think even longer term - maybe 2 to 5 years ahead. 

\begin{enumerate}
 \item Let’s say that PANL gets this moment “right” and figures out a new telework policy that allows PANL to meet new goals. What does that PANL look like in 5 years? 
 \item What is that PANL of the future good at, that you may not be good at right now?
 \item What is the culture of that PANL? How is that different from the culture now? What are the main barriers to (changing or maintaining) that envisioned culture?
 \item What do you think are the three main reasons that vision of PANL would not come to be?
 \item To end, I want to invite you to think about a common phrase in design thinking. Think about the current remote work situation. Based on that, can you complete these three phrases: ``I like... I wish... What if...''
\end{enumerate}

\section{Collaboration Technology Survey}~\label{appendix-survey}
This survey should take no more than 20 minutes to complete.

This is a survey intended to understand your use of communication, collaboration, and documentation tools at PANL. It consists of 3 basic demographic questions, followed by 12 main questions, which ask you to state your preferred tool for several basic work activities. Based on your responses, you will be asked up to three follow-up questions (two multiple choice, and one an optional free response). In the final section, there will be one open-ended question, and you will be asked about the extent to which you agree or disagree with 7 statements.

The survey is anonymous unless you would like to participate in a follow-up interview (you will be asked this at the end). Thank you so much for taking the time to complete it.

\subsection{Section 1: Demographics}

\begin{enumerate}
 \item What is your sub-organization?
 \item What is your department? 
 \item How long have you been at PANL?
 \item Are you a supervisor?
\end{enumerate}

\subsection{Section 2: Tool Use}
What is your primary tool for each of the following activities? If you do not participate in this activity, please leave it blank.

\subsubsection{1-1 Communication}
\begin{enumerate}
 \item Sharing a quick update with one teammate
 \item Sharing a quick update with a person outside of my group
 \item Asking someone a question
 \item Having a long discussion with one teammate
 \item Having a long discussion with a person outside of my group
\end{enumerate}

\subsubsection{Group Communication}
\begin{enumerate}
 \item Sharing a quick update with multiple people
 \item Starting a long discussion with multiple people
\end{enumerate}

\subsubsection{Collaboration}

\begin{enumerate}
 \item Planning a project with someone
 \item Working on a project or document synchronously with someone
 \item Soliciting feedback on a document
 \item Sharing a document with someone in my group
 \item Sharing a document with someone outside my group
\end{enumerate}

\subsubsection{Document Management}

\begin{enumerate}
 \item Store official internally facing documents 
 \item Store official externally facing documents
\end{enumerate}

\subsubsection{Follow-Up Questions}
After users select tools from the above question, the following questions will appear for each activity, only if the individual did not leave the response blank:

I use this tool because (select all that apply):
\begin{enumerate}
 \item I prefer this tool
 \item I like the features of this tool [Please specify]
 \item My team uses this tool
 \item Others I work with use this tool
 \item I know most groups use this tool
 \item My manager asked me to use this tool
 \item This tool was easily accessible on my computer
Other [Please specify]
\end{enumerate}

I would be willing to change to another tool. 
[1-7 Likert scale]\\

(Optional) Why would you be unwilling to change to another tool? [Free response]

\subsection{Section 3: Closing}

[Short response] When I have a problem with technology or tools, I seek support by...\\

For the following questions, please indicate the extent to which you agree or disagree with the provided statements.

\begin{enumerate}
 \item I feel that I’m using too many apps and technologies in my work.
 \item I believe that my work would benefit from clearer norms about using apps and technologies.
 \item (Reversed Scale) I believe that using many different apps and technologies is important to my work.
 \item I frequently switch between different devices or apps at work. 
 \item I find myself losing focus at work because my attention is divided between different devices and apps. 
 \item I find it difficult to find information on my devices or apps at work. 
 \item I feel that PANL should give me more recommendations about how best to use apps and technologies. 
\end{enumerate}

\subsection{Follow-up interview}
Would you like to participate in a follow-up interview? If so, please enter your email address.

\end{document}